\documentclass[pra,draft,twocolumn,showpacs,floatfix]{revtex4}
\usepackage{psfig}
\usepackage{bm}
\begin{document}

\title{Atom trapping and two-dimensional Bose-Einstein
  condensates in field-induced adiabatic potentials}

\author{O.~Zobay} 
\affiliation{Theoretische Quantenphysik, Technische Universit\"at 
Darmstadt,
         Hochschulstr.~4a, 64289 Darmstadt, Germany}

\author{B.~M.~Garraway}
\affiliation{Department of Physics and Astronomy, University of
         Sussex, Brighton BN1 9QH, United Kingdom}

\date{\today}

\begin{abstract}

   \quad We discuss a method to create two-dimensional traps as well
   as atomic shell, or bubble, states for a Bose-Einstein condensate
   initially prepared in a conventional magnetic trap. The scheme 
   relies on the use of time-dependent, radio frequency-induced adiabatic
   potentials. These are shown
   to form a versatile and robust tool to generate novel trapping
   potentials. Our shell states take the form of thin, highly stable
   matter-wave bubbles and can serve as stepping-stones to prepare atoms
   in highly-excited trap eigenstates or to study `collapse and revival 
   phenomena'. Their creation requires gravitational effects to be
   compensated by applying additional optical
   dipole potentials. However, in our scheme gravitation can also be
   exploited to provide a route to two-dimensional atom trapping. We
   demonstrate the loading process for such a trap and examine
   experimental conditions under which a 2D condensate may be prepared.

\end{abstract}
\maketitle

\section{Introduction}

Due to the rapid advances in experimental and theoretical atom optics,
it has now become possible to cool atoms down to extremely low
temperatures.  An important characteristic of such ultracold atoms is
their sensitivity to very weak external potentials.  In the context of
atomic Bose-Einstein condensation, this feature is particularly
attractive. The condensate can be described, to a very good degree of
approximation, in terms of a single macroscopic wave function which in
this way can be subjected to intricate probing and manipulation.
Consequently, in recent years various ingenious techniques for
handling Bose-Einstein condensates (BECs) have been developed that
exploit this sensitivity, e.g., magnetic quadrupole and TOP traps,
shallow optical dipole traps, phase-imprinting methods to create
solitons or vortices, and radio-frequency (RF) output coupling, to
name but a few prominent examples (for a review of these and other
techniques, see \cite{KetDurSta99}).  However, the development of
further tools and methods still presents a significant objective of
current work, and forms the basis for advancing research in several
main areas of BEC physics.

One such area concerns the experimental realization of low-dimensional
Bose-Einstein condensates of trapped dilute atomic gases. Under
reduced dimensionality, the condensate properties differ drastically
from the well-studied three-dimensional case and have been under
intense theoretical debate for some time \cite{2Drefs,1Drefs}.
Low-dimensional BECs are characterized by the fact that due to strong
confinement by the external trapping potential one or more motional
degrees of freedom become quantum-mechanically frozen before the
condensation point is reached. Major obstacles, that had been
confronting the preparation of one- or two-dimensional BECs for a long
time, concerned the construction of suitable traps---e.g., very high
field gradients would be required for magnetic traps---as well as the
development of efficient loading procedures.  Only recently has it
become possible to overcome these difficulties, and the first
realizations of atomic BECs in two dimensions have been reported
\cite{GorVogLea01,BurCatFor02,MorCriMul02}. In these experiments, the
dimensionality is reduced by exposing the atoms to steep optical
potentials. In Ref.\ \cite{GorVogLea01}, single one- or
two-dimensional BECs are prepared, whereas in the experiments of
Refs.\ \cite{BurCatFor02} and \cite{MorCriMul02} the use of optical
lattices led to the creation of arrays of two-dimensional condensates.
However, in addition to these works, various other interesting ideas
regarding the manufacture of two-dimensional traps have been put
forward over the last few years. In these proposals, trapping is
provided either by optical \cite{Opt} or by magnetic
\cite{HinBosHug98} means, whereas loading is accomplished through
optical pumping.

The purpose of the present paper is twofold. On the
one hand we wish to promote field-induced adiabatic potentials as a new
versatile tool to manipulate ultracold atoms and, in particular,
Bose-Einstein condensates. On the other hand, we show that they offer
a novel route to the creation of two-dimensional traps for BECs.
An initial outline of our results was given in Ref.~\cite{ZobGar00a}.
Adiabatic potentials arise whenever two or more internal atomic states,
that experience different potentials for the atomic center-of-mass
motion, are coupled by a strong resonant external field. The atomic
motion is then no longer dominated by the different `bare' potentials but
is usefully described in terms of the adiabatic potentials, that arise from
the diagonalization of the bare potentials and the couplings at each
spatial point. Adiabatic potentials have been in use for some time as the
underlying mechanism for evaporative cooling. However, to our knowledge,
little attention has been paid to the fact that they offer a lot of
further possibilities to control quantum-mechanical atomic motion (one 
fairly recent
application is described in \cite{BloKoeGre00}).
In this paper we will give some examples of the application of adiabatic
potentials as a tool for manipulating matter waves, and we
hope to stimulate further research in this direction.

After giving a qualitative introduction to adiabatic potentials in
Sec.\ II, we continue in Sec.\ III to show how they can be employed to
create so-called `matter-wave bubbles' from a BEC initially trapped in
the ground state of a harmonic potential. In the bubble state the wave
function is localized around the surface of a sphere so that the
matter density forms a spherical shell or bubble. We give a detailed
account of the preparation of bubble states and their decay rates
induced by nonadiabatic leakage.  In particular, we address the
question of how to compensate for gravitational effects in the
laboratory which otherwise would impede the creation process.  We
expect matter-wave bubbles to have interesting applications some of
which are investigated in Sec.\ IV. There it is shown that they can be
used as stepping-stones to prepare atoms in highly excited eigenstates
of the bare harmonic trapping potential and to create new types of
non-linear eigenstates of BECs. Furthermore, we examine collapse and
revival effects which are found when the external coupling is switched
off.  By also switching off the trapping potential we can observe the
free bubble expansion.

In Sec.\ V we turn to the investigation of gravitational influences on
the bubble preparation process. At first sight gravity is seen to be
detrimental, but it is shown that it can be exploited to obtain a
novel scheme for the creation of two-dimensional BECs. Gravity will
cause the atoms in the shell potential to pool at the bottom of the
trap so that the condensate forms a flat disk.  In Sec.\ V the initial
transfer of a BEC into such a disk state in the course of switching on
the external field is examined numerically.  In particular, we show
that if the field parameters are changed appropriately the disk radius
is steadily increased whereas its width keeps shrinking due to
enhanced confinement. This suggests that the continuation of this
process could ultimately yield a two-dimensional trap for the BEC,
although the numerical study of the approach to this limit is not
currently feasible.  In Sec.\ VI we further pursue this idea by giving
some general qualitative, and semi-quantitative, estimates and
arguments about the conditions necessary to reach the 2D regime. They
indicate the feasibility of our approach by showing that the
requirements for the applied fields, preservation time of the
condensate etc.\ are demanding but still within the reach of currently
available technology.  The paper ends with brief conclusions and
outlook given in Sec.\ VII.

The most important difference to the previous proposals of Refs.\ 
\cite{Opt,HinBosHug98} for creating 2D traps is the fact that our
method relies on adiabatically deforming a conventional magnetic trap
and does not require incoherent processes, e.g., optical pumping, for
loading. This would allow for working with extremely cold, dense, and,
possibly, coherent atomic ensembles throughout the whole process. In
comparison to the experiments
\cite{GorVogLea01,BurCatFor02,MorCriMul02}, our scheme does not make
use of optical potentials, but a combination of magnetic and RF
fields.  It thus avoids potential difficulties with spontaneous
emission in very steep optical traps that require high laser
intensities. In contrast to Refs.\ \cite{BurCatFor02,MorCriMul02}, our
proposal produces a single condensate with a large number of atoms,
similar to the experiment of Ref.~\cite{GorVogLea01}. However, the
trapping frequencies obtainable in the RF-scheme might be
significantly higher trapping frequencies than the ones reported in
the latter work.

\section{Adiabatic potentials}

The basic principles of adiabatic
potentials can be understood from examining a quantum-mechanical
two-state particle that propagates in the vicinity of a linear potential
crossing. In the interaction picture, with respect to the coupling field
that gives rise to the crossing,
the Schr\"odinger equation for this system is written as follows:
\begin{eqnarray}\label{lin}
i\dot\phi_1&=&\left(-\frac 1 2 \frac{\partial^2}{\partial
r^2}+Cr\right)\phi_1
+\Omega \phi_2, \nonumber \\
i\dot\phi_2&=&\left(-\frac 1 2 \frac{\partial^2}{\partial
r^2}-Cr\right)\phi_2
+\Omega \phi_1.
\end{eqnarray}
We can transform this equation to a basis that diagonalizes the
potentials $\pm Cr$ and the coupling $\Omega$ at each point, i.e., to the
basis of the spatially dependent `dressed eigenstates.' In this basis
the Schr\"odinger equation has the form \cite{LevJohBer69}
\begin{eqnarray}\label{dres}
i\dot\phi_+&=&\left[-\frac 1 2 \frac{\partial^2}{\partial r^2}+V_+(r)
+V_{kin}(r)
\right]\phi_+ + V_c(r,\partial_r) \phi_-\nonumber \\
i\dot\phi_-&=&\left[-\frac 1 2 \frac{\partial^2}{\partial r^2}+V_-(r)
+V_{kin}(r)
\right]\phi_- - V_c(r,\partial_r) \phi_+.
\end{eqnarray}
The potentials $V_{\pm}(r)$ are the adiabatic potentials and are given
by
\begin{equation}\label{adpot0}
V_{\pm}(r,t)=\pm \sqrt{(Cr)^2+\Omega^2}.
\end{equation}
They arise from the pointwise diagonalization of the $2\times2$-matrix
formed by $\pm Cr$ and $\Omega$. The terms
\begin{equation}
V_{kin}(r)=\frac{C^2}{8\Omega^2[1+(Cr/\Omega)^2]^2}
\end{equation}
and
\begin{equation}
V_c(r,\partial_r)=\frac{C}{2\Omega[1+(Cr/\Omega)^2]}
\left[\frac{C^2r}{\Omega^2[1+(Cr/\Omega)^2]}-\frac{\partial}{\partial r}
\right]
\end{equation}
stem from the nonlocal character of the kinetic energy term.

We see that $V_+(r)+V_{kin}(r)$ is a binding potential. If the
coupling term $V_c(r,\partial_r)$ were not present, a wave packet
prepared in the state `$+$' would remain trapped forever in the
potential $V_+(r)+V_{kin}(r)$.  However, if $\Omega$ is small the
coupling $V_c$ is dominant. In this case any wave packet quickly
leaves the crossing region around $r\simeq 0$, and the description
given by Eq.\ (\ref{dres}) is not useful. Nevertheless, if $\Omega$ is
increased, the kinetic coupling $V_c$ (as well as $V_{kin}$) rapidly
becomes small\footnote{It should be noted that although the magnitude
of $V_c$ is reduced for growing $\Omega$, the effective interaction
range $\Delta r \simeq \Omega/C$ is increased. However, the size of
the ground state of $V_+$ scales with $\Omega^{1/4}$, i.e., it only
experiences the diminished magnitude of $V_c$.}  and the motion of
the wave packet is more and more determined by the adiabatic
potentials $V_{\pm}$. In fact, it is shown in Sec.\ III.B that the
lifetime of a wave packet prepared in the internal state `$+$', i.e.,
the time it takes the wave packet to transfer to the state `$-$' and
leave the crossing region, increases exponentially with $\Omega$.

At first sight, it seems counterintuitive that it is possible to trap
a particle, even with strong coupling, between two potentials in a
region where the particle is not stable.  We can draw an analogy with
a particle moving in a magnetic quadrupole field. Each of its bare
Zeeman substates (with bare meaning having fixed spatial orientation)
individually would experience the field as unstable. However, if the
particle moves slowly enough, couplings between the states are induced
so that the particle's orientation with respect to the local direction
of the total magnetic field is preserved. In this way, trapping can
ensue for weak-field seeking states.  In the case considered here, a
slow particle tends to remain in the same dressed eigenstate and its
motion is governed by the adiabatic potentials.

\section{Creation of matter-wave bubbles}

\subsection{Basic approach}

To work out the basic ideas of our approach we first discuss its
realization in the absence of gravity. As a result, we obtain a scheme
to produce thin, highly stable matter-wave bubbles or shells
in which the trapped atoms are localized around the surface of
a sphere.

\begin{figure}
\centerline{\psfig{figure=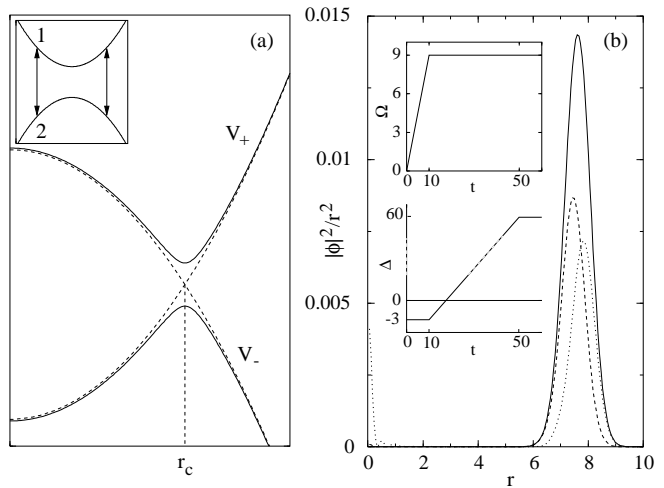,width=8.6cm,clip=}}
\caption[f1]{(a) Schematic of field-induced adiabatic potentials 
$V_{\pm}$ for $\Delta>0$. Dashed curves show the bare potentials
crossing at $r_c$.  Inset: bare potentials showing resonance at
$r_c$.  (b) Bubble, or shell state as obtained by the preparation scheme
discussed in the text with (in scaled units) $\Delta_{final}=60.0$ and
$\Omega_{final}=9.0$ (see inset). Full curve: atomic density
$|\phi_+|^2/r^2$ in the adiabatic state $+$; dotted and dashed
curves show $|\phi_1|^2/r^2$ and $|\phi_2|^2/r^2$, respectively.
Nonlinear effects are not included here.
}
\label{fig1}
\end{figure}

The starting point for our method is a coherent sample of atoms
produced, e.g., by Bose-Einstein condensation and trapped in the
ground state of a harmonic magnetic potential. The preparation scheme
then proceeds by applying a sequence of radio frequency fields that
couple the initial internal atomic state, a weak-field seeking Zeeman
sublevel, to a second hyperfine ground state (multi-level excitation
schemes can be considered as well, see Sec.\ V).  One interesting
aspect of the preparation process is that it may appear to be
counterintuitive as the second state is an untrapped, high-field
seeking state [see inset, Fig.\ \ref{fig1}(a)].  As this technique
also forms the basis of evaporative cooling, one may be led to expect
that this procedure will inevitably cause a rapid depletion of the
trapped atomic population. Nevertheless, as was anticipated in Sec.\ 
II and is also shown below, if the fields are controlled in an
appropriate way one may also obtain a very different effect.

To model the creation of our bubble states
we study the coherent time evolution of a condensate
initially prepared in a hyperfine state $|1\rangle$ in the
ground state of a spherically symmetric magnetic
trap.  When $t>0$ external fields are applied that induce a
coupling of state $|1\rangle$ to a Zeeman sublevel $|2\rangle$
whose magnetic moment is supposed to be equal in magnitude but
opposite in sign.  Such a coupling may be realized, e.g., as a
transition between the hyperfine ground states
$|F=1,m_F=-1\rangle$ and $|F=2,m_F=-1\rangle$ in $^{87}$Rb
\cite{JILArefs}.  The field-induced coupling strength is denoted
as $\Omega$ and is spatially independent to a good degree of
approximation. In the following model we assume a radial $l=0$
form of the wave function components $\psi_i({\bf
r})=\phi_i(r)/\sqrt{4 \pi}r$.  Then working in an interaction
picture with respect to the applied fields, the time development
is determined by the radial Gross-Pitaevskii equation
\begin{eqnarray}\label{GPE}
 i\dot\phi_1&=&\left(-\frac 1 2 \frac{\partial^2}{\partial r^2}+\frac 
{r^2}2
 -\frac{\Delta(t)}2\right)\phi_1+\Omega(t) \phi_2 +\nonumber \\
 & & N\left(U_{11} |\phi_1|^2 +U_{12} |\phi_2|^2\right)\frac{\phi_1}{r^2},
 \nonumber\\
 i\dot\phi_2&=&\left(-\frac 1 2 \frac{\partial^2}{\partial r^2}-\frac {r^2}
2
 +\frac{\Delta(t)}2\right)\phi_2+\Omega(t) \phi_1 +\nonumber \\
 & & N\left(U_{12} |\phi_1|^2 +U_{22} |\phi_2|^2\right)\frac{\phi_2}{r^2}.
\end{eqnarray}
In this equation time is scaled to the trap inverse oscillator
frequency $\omega^{-1}$, length is scaled to the harmonic oscillator length
scale $a_{ho}$, and $\Omega(t)$ and $\Delta(t)$ are scaled to
$\hbar\omega$.  The effective detuning is defined by
$\Delta(t)=[\hbar\omega_{f}-\Delta E(0)]/\hbar\omega$, where
$\Delta E(0)$ is the energy difference between the two hyperfine
states at the origin (trap centre) and $\omega_f$ the frequency of the 
applied
field.  The nonlinearity parameters are given by
$U_{ij}=a_{ij}/a_{ho}$ with $a_{ij}$ the scattering lengths for
intra- and inter-species collisions.
The states are normalized according to $\int_0^{\infty}
dr(|\phi_1|^2+|\phi_2|^2)=1$. The total number of atoms is denoted $N$.

The strategy that we pursue in our engineering scheme is to
control the condensate by slowly changing field-induced adiabatic
(or dressed) potentials. These potentials, which are defined as
the spatially dependent eigenvalues of the potentials and
couplings in Eqs.\ (\ref{GPE}), are given by [cf.\ Eq.\ (\ref{adpot0})]
\begin{equation}
  V_{\pm}(r,t)=\pm \sqrt{[r^2-\Delta(t)]^2/4+\Omega^2(t)}
\end{equation}
and are depicted for $\Delta>0$ in Fig.\ \ref{fig1}(a). The potential
$V_-$ actually gives rise to the evaporative cooling effect in the
usual arrangement. In that case one applies a field with an effective
detuning which is large compared to the mean particle energy. The
atoms then move in the potential $V_-$, and the ones that reach its
maximum, from the left, with sufficiently slow velocity, go over the
top and get expelled from the trap. However, we will show that the
atoms can also be prepared in the lowest-energy eigenstate (or, more
exactly, resonance) $|0\rangle\equiv |0;\Omega,\Delta\rangle$ of the
potential $V_+$. This quasi-bound or `trapping' state $|0\rangle$ will
realize the spherical shell state or matter-wave bubble (see Fig.\ 
\ref{fig1}(b) and Ref.\ \cite{ZobGar00}).  The state will be localized
around the crossing of the two bare potentials at $r_c=\sqrt{\Delta}$
and have a width of $\Delta r=(\Omega/\Delta)^{1/4}$, provided a
harmonic expansion around the potential minimum is justified. The
state is a genuine superposition of the internal states $|1\rangle$
and $|2\rangle$.

\subsection{Lifetime of bubbles}

\begin{figure}
\centerline{\psfig{figure=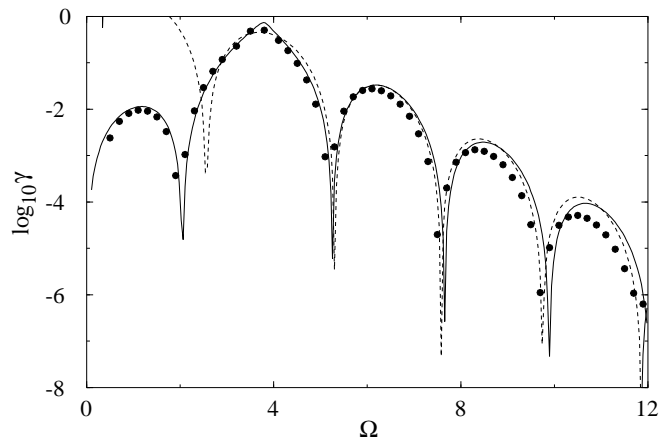,width=8.6cm,clip=}}
\caption[f2]{
 Evaluation of decay rates $\gamma(\Omega)$ for bubble states
 according to Eqs.\ (\protect\ref{eqE}) (full curve) and
 (\protect\ref{appgamma}) (dashed) at $\Delta=61.0$, using the
 approximations of Ref.\ \protect\cite{BanChi70}.  Circles:
 numerical values determined from Eq.\ (\protect\ref{GPE}).}
\label{fig2}
\end{figure}

Before considering the bubble preparation process in detail one
question immediately arises, i.e., the stability of the system once it
is prepared in the state $|0\rangle$. At first it is not obvious that
atoms may remain trapped for a substantial time at the point of
maximum effective coupling between states $|1\rangle$ and $|2\rangle$.
As we have discussed in Sec.\ II, it becomes more plausible if one
transforms Eqs.\ (\ref{GPE}) to the dressed state basis, i.e., the
basis that diagonalizes the bare potentials and the coupling at each
point $r$. In this picture the two wave function components appear
coupled by kinetic terms whose significance is rapidly diminished when
$\Omega$ is increased.  If non-linear interactions can be neglected,
the decay rate $\gamma$ of the trapping state $|0\rangle$ can be
determined with the help of semi-classical methods developed in
connection with molecular predissociation \cite{BanChi70,Chi74}. As we
show in the Appendix, from these techniques it follows that
$\gamma=-2\mbox{Im}\,E_0$, where the complex ground state energy $E_0$
is determined as a solution of
\begin{equation}\label{eqE}
[e^{2\pi \delta(E)}-1]\cos\Phi(E) e^{-i[\beta(E)-\Phi(E)]}+
\cos\beta(E)=0
\end{equation}
with $\beta(E)=\pi(2E+\Delta-1)/4$ and the parameters $\delta(E)$
and $\Phi(E)$ characterizing the scattering matrix of the linearized
potential crossing problem. For these quantities there are 
several analytical approximations
in the literature \cite{ZhuNak93}; following, e.g.,
Ref.\ \cite{BanChi70} one can put $\delta(E)=1/8ab$ and
 $
 \Phi(E)=2b^3/3a+\mbox{arg}\Gamma[i\delta(E)]+\delta(E)\ln
 [\delta(E)]-2\delta(E)\ln(b/a)+\pi/4
 $ 
with $a^2=\Delta/(8\Omega^3)$ and $b^2=E/\Omega$. For large enough $\Omega$
one obtains 
\begin{equation}\label{appgamma}
 \gamma=\frac{2\cos^2 \beta(E) }{\{\exp[2\pi\delta(E)]-1\}(\partial
 \Phi/\partial E)}
\end{equation}
where all quantities have to be evaluated at $\mbox{Re}E_0\simeq
\Omega+\sqrt{\Delta/4\Omega}$.  The comparison in Fig.\ \ref{fig2} between
the predictions of Eqs.\ (\ref{eqE}) and (\ref{appgamma}) and the direct
numerical determination of decay rates from Eqs.\ (\ref{GPE}) indicates the
validity of these approximations. To obtain the numerical decay rates we
first generated the bubble state at a required $(\Delta,\Omega)$
by performing the preparation process described in Sec.\ III.C.
We then monitored the decay of the norm of the wavefunction in the
internal state `$+$' as a function of time. After an initial transient,
this decay was exponential to a very good degree of approximation.

Equation (\ref{appgamma}) yields two
important insights: firstly, the decay is exponentially suppressed with
growing $\Omega$.
In the limit of $\gamma \ll 1$, this exponential
suppression may be approximated as
\begin{equation} \label{gamma2}
\gamma_{ex}\sim 2\exp(-\pi\Omega^{3/2} /\sqrt{2}\Delta^{1/2}).
\end{equation}
Secondly, for $\mbox{Re}E_0=2k+ 3/ 2 
- \Delta/2$, with integer $k$, the decay rates become very small. In
these cases the state $|0\rangle$ is in resonance with an eigenstate of the
bare harmonic trapping potential. This stabilization effect may be used to
obtain extremely long-lived states already for moderate coupling strengths.

\subsection{Preparation}

Having established the existence of long-lived spherical bubble states
in adiabatic potentials, we now turn to the details of their
preparation. The first phase is to let the bare harmonic potential of
state~1 evolve smoothly into the dressed potential $V_+$ and the
second phase is to expand the bubble outwards.  Both phases are done
so that the wave function is kept at all times in the instantaneous
ground state $|0;\Omega(t),\Delta(t)\rangle$. In the example of Fig.\ 
\ref{fig1}(b) (in which non-linear effects have been neglected) we
achieve the first phase by increasing $\Omega$ to the final (scaled)
value of 9 at fixed {\it negative} detuning $\Delta=-3$ (see Fig.\ 
\ref{fig1}(b), inset). Then the second phase of bubble expansion is
achieved by maintaining the intensity, but gradually ramping up the
detuning.  In this way less than 4\% of the initial population was
lost before reaching the final state shown in Fig.\ \ref{fig1}(b). The
particular merit of this approach is that there is no need to follow a
precisely prescribed path as the method relies on adiabatic following
(in a similar way to the wave packet guidance in APLIP \cite{kas}).
Any field sequence that guides the system sufficiently slowly through
states of long lifetime is suitable.

If we make our atomic bubbles from an initial BEC, we should also
study the effects of the non-linear interactions introduced in
Eq.\ (\ref{GPE}).  The dressed-state description is readily
generalized if $U_{11}\simeq U_{22}\simeq U_{12} = U$ (such a
situation is realized, e.g., in the Na $F=1$ hyperfine multiplet
\cite{KetDurSta99}). Under these circumstances we can still
define an adiabatic wave function $\phi_+$ that evolves, to a
good degree of approximation, according to the Gross-Pitaevskii
equation
\begin{equation}
i\dot\phi_+=\left[-\frac 1 2 \frac{\partial^2}{\partial r^2}+V_+(r,t)
+NU\frac{|\phi_+|^2}{r^2}\right]\phi_+.
\end{equation}
In particular, for a sufficiently strong nonlinearity $U$, the atomic
density can be described by the Thomas-Fermi ansatz
\begin{equation}
 |\phi_+|^2=[\mu - V_+(r,t)]/NU
\end{equation}
with the chemical potential $\mu$ determined by the normalization 
condition. Numerical studies show that shell, or bubble, states can then 
be manufactured in the
same way as before. Once prepared, their decay is now non-exponential, but
the rate is of the same order of magnitude as the corresponding
system in the linear case. If the atomic cross- and self-interactions
differ significantly, numerical simulations indicate that the creation
of bubble states becomes more difficult and their lifetime is reduced.

\subsection{Bubbles with gravitational field compensation}
\label{subsec:bubbgrav}

So far, we have neglected the influence of gravity. Its effect on the
bubbles is expected to be detrimental as it will cause the atoms to
pool at the bottom of the shell potential. In this subsection, we first
give a simple estimate of how strong the gravitational influence
must be to cause significant deviations from a homogeneous density
distribution in the bubble state. We then propose a method to compensate
for gravitation in the laboratory with the help of optical potentials.

When the bubble width is small compared to the radius, the
radial dependence of the adiabatic potential is approximately
 harmonic with a minimum at the bubble
radius $r_0=\sqrt{\Delta}$ and an angular frequency
$\omega_0=\sqrt{\Delta/\Omega}$ \cite {ZobGar00}.
Thus, we can approximate the Hamiltonian for atoms in the trapping
internal state `$+$' as 
\begin{equation}
H=-\frac{\nabla^2}2+\frac 1 2 \omega_0 (r-r_0)^2+Gr\cos \theta,
\end{equation}
with the radial and polar coordinates  $r,\theta$, and
the scaled gravitational acceleration $G=g\sqrt{m/\hbar \omega_z^3}$ 
(with $g\simeq 9.81$ms$^{-2}$).
We now make the ansatz
\begin{equation}\label{ansatz}
\psi_{tr}(r,\theta)=\frac {1}{\sqrt{\cal N}} \frac{\exp[-(r-r_0)^2/2 
\sigma^2]}r f(\cos\theta)
\end{equation}
for the ground state of this Hamiltonian, where
$\sigma=1/\sqrt{\omega_0}$. This ansatz implies that, radially, the wave
function is always in the ground state, i.e.\ the influence of gravity is
manifest only in the polar envelope $f(\cos\theta)$.
For present purposes
it is sufficient to choose a very simple form for $f$, e.g.,
$f(\cos\theta)=a\cos\theta +b$. The normalization factor in Eq.\
(\ref{ansatz}) is then given by ${\cal N}=\sqrt{\pi}(2a^2/3+b^2)/\sigma$.
Furthermore, a variational calculation shows that $a$ and $b$ are
related by $a/b=Q-\sqrt{Q^2+3/2}$ with $Q=3/4Gr_0^3$. A
significant influence of gravity is certainly present when $f(\cos 0)=0$,
i.e., $a/b=-1$, as in this case the bubble has `opened up.' This is 
realized
for $Q\simeq 0.17$ or, in unscaled coordinates,
\begin{equation} \label{maxr0}
r_0\simeq (3\hbar^2/gm^2)^{1/3}
\end{equation}
which is about $5.5\times 10^{-7}$m for Rb87. In the presence of
gravity, bubbles can therefore only be observed in spherical traps with
trapping frequencies large compared to 400Hz. The estimate (\ref{maxr0}) 
can
be interpreted as a balance condition between kinetic and gravitational
energy which are of the order $\hbar^2/2mr_0^2$ and $mgr_0$, respectively.

\begin{figure}
\centerline{\psfig{figure=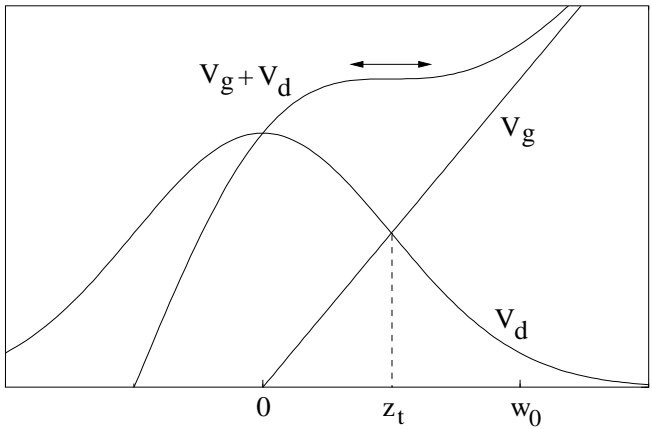,width=8.6cm,clip=}}
\caption{Schematic showing the local compensation of gravity with the
help of an optical potential. Shown are dipole potential $V_d$, the
gravitational potential $V_g$, and their sum. The arrow indicates the
flat region of the combined potential in which a bubble might be created.}
\label{figopt}
\end{figure}

The detrimental influence of gravity can, at least in principle, be
compensated for by exposing the trapped atoms to an additional optical
dipole potential. In a typical Gaussian beam configuration this
potential, which acts on all hyperfine sublevels in the same way, is given 
by
\cite{KetDurSta99}
\begin{equation}\label{optpot}
V_d(x,\varrho)=\frac{V_0}{1+(x/x_R)^2}\exp\left\{
-\frac{2\varrho^2}{w_0^2[1+(x/x_R)^2]}\right\}
\end{equation}
with $\varrho=\sqrt{y^2+z^2}$. The Rayleigh length $x_R$ and the
beam waist radius $w_0$ are related by $x_R=\pi w_0^2/\lambda_{op}$ with
$\lambda_{op}$ the wavelength of the optical field. Finally, in unscaled
units, $V_0=3\Gamma c^2P/\Delta w_0^2\omega_0^3$ where $\omega_0$
denotes the resonance frequency, $\Gamma$ the spontaneous decay rate
of the excited state, $\Delta$ the detuning between laser and the atomic
transition and $P$ the applied laser power. The idea is now to adjust
the trap depth $V_0$ such that at the turning point $z_t$ of the optical
potential along the line $x=y=0$, which is defined by $\partial^2 V_d
(z_t)/\partial z^2=0$, the slope $\partial V_d(z_t)/\partial z=-mg$.
In this way, the combined optical and gravitational potential is almost
constant around $z_t$, the lowest-order corrections being cubic in $z-z_t$
(see Fig.\ \ref{figopt}).
The magnetic trap has then to be placed inside this area.
 From Eq.\ (\ref{optpot}) it follows that, as $z_t=w_0/2$, we have to 
choose
$V_0=\exp(1/2)w_0mg/2$ to fulfill the above condition.
If the condensate is in the Thomas-Fermi regime, the $z$ extension of
the volume within which a bubble could be created is somewhat less than
$(w_0^2\mu/mg)^{1/3}$ with $\mu$ the BEC chemical potential [for which an
estimate is given in Eq.\ (\ref{chempot})]. This follows from
stipulating that the variation
of the combined optical and gravitational potential over this volume
should be less than $\mu$.

We have performed numerical simulations to demonstrate that
this approach actually offers the possibility to overcome the influence of
gravity. To this end, we have studied the preparation process using a
two-dimensional version of Eq.\ (\ref{GPE2}) including the optical 
potential. 
Equation (\ref{GPE2}) describes time evolution in the effective
adiabatic potential of the Rb87 $F=2$ multiplet. As an example we have 
shown
that in an anisotropic trap with $\nu_z=\omega_z/2\pi=220$Hz,
$\nu_x=30$Hz, bubble states for a BEC
corresponding to an atom number of 10$^5$ in three dimensions can be
produced at $\Delta=6.6$kHz, $\Omega=2.6$kHz, $V_0=-129$kHz, and 
$w_0=73\mu$m.
For the preparation, $\Omega$ is switched
on within $\Delta t=29$ms at $\Delta=-1.1$kHz, and then $\Delta$ is 
increased to
its final value within $\Delta t=200$ms. The bubble has radial and
axial diameters of 15$\mu$m and 110$\mu$m, respectively.
The necessary laser power is estimated
at around 1W. The large axial extension shows that in this direction the 
variation
of the combined optical and gravitational potential is almost negligible.
To find an optimized shape for given $\Omega$ and $\Delta$,
it is advisable to experiment by slightly varying the
laser power and the position of the magnetic trap center.
Nevertheless, the preparation scheme is robust in the sense that
(for the configuration described above)
one still obtains a bubble if 
the location of the magnetic trap (in $z$) relative to the
dipole potential is changed by about $\pm$10\%, and the laser
intensity by about $\pm$2.5\%. However, {\it during} preparation 
these parameters have to be stabilized very well (to within a fraction 
of the indicated range) to avoid excitation of the bubble.

\subsection{Bubbles and trap anisotropies}

At this point it should be mentioned that---at least in the absence of
non-linear interactions---the
preparation of bubbles is impeded by another problem besides gravity,
namely, the unavoidable trap anisotropies. Imagine that in the bare
magnetic trap $\omega_x=\beta\omega_z$, but we keep $\beta\approx 1$.
Then the effective trapping
frequency along the $x$ axis is given by $\beta\sqrt{\Delta/\Omega}$ 
(scaling
now with respect to $\omega_z$). We can model our system Hamiltonian as
\begin{equation}
H=-\frac{\nabla^2}2+\frac 1 2 \omega_0^2(\theta) (r-r_0)^2,
\end{equation}
where $\omega_0(\theta)$ appropriately interpolates between
$\sqrt{\Delta/\Omega}$ and $\beta\sqrt{\Delta/\Omega}$. The ground state
is approximated as
\begin{equation}\label{ansatz1}
\bar\psi_{tr}=\frac 1{\sqrt{\bar{\cal N}}} \frac{\exp[-(r-r_0)^2/2
\sigma(\theta)^2]}r
\bar f(\cos\theta)
\end{equation}
with $\sigma(\theta)=1/\sqrt{\omega_0(\theta)}$. We again could
determine $\bar f(\cos\theta)$ by minimizing the energy functional.
However, following the discussion of the previous paragraph
we can argue that variations
in the angular kinetic energy carry a cost of the order of $1/\Delta$
whereas the difference in `polar potential energy' is given by $|\beta^2-1|
\Delta/\Omega$.
Therefore, we expect trap anisotropies to
have a significant influence as soon as
\begin{equation}
\left|\left(\frac{\omega_x}{\omega_z}\right)^2-1\right|\frac{\Delta}{\Omega}
\,\, \agt \,\, \frac 1{\Delta}.
\label{frequencies}
\end{equation}
As typically $\Delta \gg 1$ and $\Delta/\Omega \gg 1$, the trap frequencies
have to be adjusted very carefully in order to enable the generation
of bubbles. In the case of Fig.\ 2, for example, the estimate
(\ref{frequencies}) indicates that $|\omega_x/\omega_z -1|$ should
be of the order of 0.1\%.
Fortunately, the
problem can be alleviated to some extent by making use of the atomic
interactions. In the Thomas-Fermi limit, the chemical potential of a
thin bubble is estimated to be
\begin{equation}\label{chempot}
\mu=\left(\frac{3}{4\sqrt{2}}\frac{UN}{\sqrt{\Omega\Delta}}
\right)^{2/3}
\end{equation}
with $U=a_{scat}/a_{ho}$ as the scaled nonlinearity coefficient and
where $N$ is the number of atoms in the bubble. Thereby, $\mu$ is measured
with respect to the bottom of $V_+(r)$. We expect that only if the
non-linear interaction energy---which is of the order of $\mu$---is in
sufficient excess of the effective potential energy, i.e., if
\begin{equation}
N \gg \frac{4\sqrt{2}}{3U}|\beta^2-1|^{3/2}\frac{\Delta^2}{\Omega},
\end{equation}
then
a bubble of approximately constant density can be formed. Thus, even if
gravity is compensated for, bubbles can probably only be produced in the
presence of strong non-linear interactions. On the other hand, the
numerical simulations of the
previous subsection have shown that under such conditions a
preparation may indeed be possible.

\section{Applications of bubbles}

\subsection{Production of excited harmonic oscillator states}

\begin{figure}
\centerline{\psfig{figure=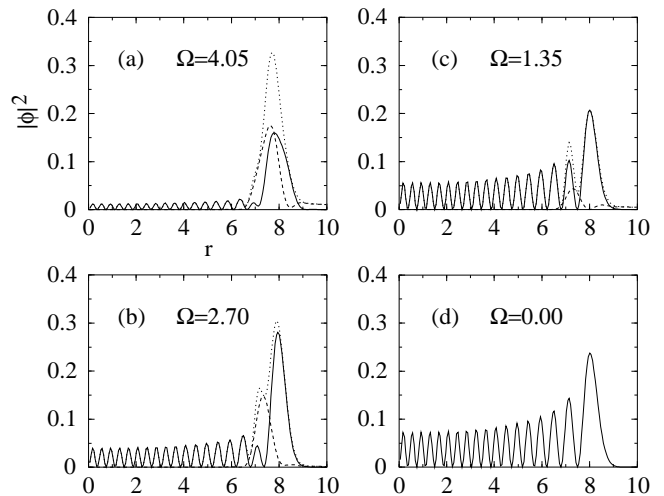,width=8.6cm,clip=}}
\caption{Resonance states of Eqs.\ (\protect\ref{GPE}) at
$\Delta= 60$ and various values of $\Omega$ (nonlinearities not taken
into account). The wave functions are determined
numerically by slowly decreasing the value of $\Omega$ in Eqs.\
(\protect\ref{GPE}) after initially preparing a bubble state. Full
curves: $|\phi_1|^2$, dashed: $|\phi_2|^2$, dotted: sum of both.
The displayed wave functions are normalized to one.
}
\label{fig4}
\end{figure}

As a first application of matter-wave bubbles we now discuss the 
preparation
of excited harmonic oscillator states in the absence of non-linear
interactions (i.e., for sufficiently dilute samples). If, after creating
a bubble state, the coupling strength is slowly 
{\it reduced at fixed} $\Delta$
the system will again evolve through a sequence of instantaneous
eigenstates. This time, however, the eigenstate will not significantly
change its energy relative to the minimum of the bare
trapping potential. Qualitatively speaking, the wave function
gradually moves out from the crossing region where it is being
trapped and begins to experience more strongly the presence of
the bare harmonic potential. This process is accompanied by a loss
of atoms which end up on the repulsive potential after leaving the
crossing region. At $\Omega=0$ the remaining bound wave function
will have reached an excited harmonic oscillator eigenstate
which, in the spherical harmonic case, is characterized by having
$l=0$. As mentioned before, the energy of this eigenstate is
approximately equal to the energy of the initial bubble state. To
give an example, if we start from the state shown in Fig.\
\ref{fig1}(b), and the coupling strength is ramped down within a
time interval of $\Delta t=16$, we arrive at an eigenstate of
energy 35.5 with respect to the minimum of the harmonic
potential \cite{ZobGar00}. 
A sequence of intermediate states appearing in the course of
this process is shown in Fig.\ \ref{fig4}.  After completion, the admixture
of other eigenstates is less than 2\%, and the population is 36\% of the
initial population in the harmonic trap ground state.

A few remarks should be made about this preparation scheme: (i)
It has to be performed `quasi-adiabatically', i.e., slow enough to
avoid substantial excitation of other eigenstates, but
sufficiently fast to reduce losses as much as possible. (ii)
Although the system effectively ends up with the same Hamiltonian
as in the very beginning (coupling switched off) and the time
evolution is performed quasi-adiabatically, the final state is
very different from the initial one. This is because the
Hamiltonian depends on two external parameters $\Omega$ and $\Delta$,
and $\Omega=0$ represents a singularity in the sense that the value
of $\Delta$ becomes irrelevant. 
(iii) The oscillator eigenstates can
in turn be used as intermediate states to produce radially
excited bubble states. To this end, one simply switches the field
coupling on again with a reduced detuning.

\subsection{Nonlinear eigenstates}

\begin{figure}
\centerline{\psfig{figure=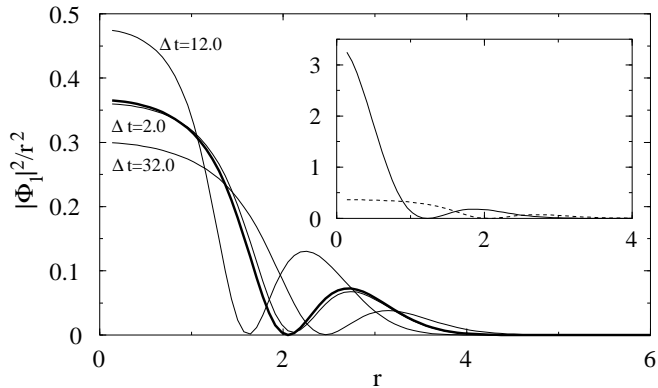,width=8.6cm,clip=}}
\caption[f3]{Condensate density $|\phi_1|^2/r^2$ at times $\Delta t=2$, 12,
and 32 after completion of the preparation scheme.
Preparation proceeds from
the bubble state at $\Omega=5.6$ and $\Delta=5.4$ by reducing $\Omega$ to 0
within $\Delta t=70$. The bold curve shows the exact `non-linear mode' for
the final value $NU_{11}=17.2$. The inset compares
this state (dashed) to the corresponding eigenstate in the absence of
nonlinearities (full curve).}
\label{fig3}
\end{figure}

The bubble states can be regarded as the ground states of a
specifically tailored potential. Recently, however, the study of
macroscopically excited states of BECs has received much attention
\cite{VorSolrefs,FedMurShl99} and the manufacture of harmonic
oscillator eigenstates, as presented above, indicates a way to prepare
a new class of such excited states. This class can be thought of as
the non-linear generalization of the eigenstates of linear systems;
for a one-dimensional model some properties of such `non-linear modes'
were examined in Ref.\ \cite{KivAleTur99} without discussing ways for
their actual preparation. For highly excited states, non-linear
effects are expected to play a minor role, in general, because of the
reduced density. We thus focus here on the first excited non-linear
$l=0$ mode which is characterized by one radial node in the wave
function. In the absence of non-linear interactions the preparation
scheme of Sec.\ IV.A works equally well for low- and high-lying
eigenstates, so we have numerically applied the same approach, with
suitably chosen values for $\Omega$ and $\Delta$, to the full
Gross-Pitaevskii equation.  Our studies show that the scheme is still
applicable, though the atomic interactions cause detrimental effects:
shortly before reaching the final value of $\Omega=0$ one encounters a
very strong loss of atoms, and, because of this, the process has to be
performed rather quickly resulting in an appreciable excitation of the
final wave function.  Furthermore, the final detuning $\Delta$ has to
be selected more carefully (to within $\pm 0.1 \omega$) to optimize
the number of atoms that remain trapped. In spite of these
difficulties, satisfactory results can be obtained for final
nonlinearity parameters $NU_{11}$ up to the order of 10. As an
example, Fig.\ \ref{fig3} shows the density distribution
$|\phi_1(r)|^2/r^2$ at various instances after completion of the
preparation process (for which $U_{11}=U_{22}=1.08U_{12}$ was assumed,
as in $^{23}$Na). The breathing seen in the wave function indicates
that there is additional radial excitation.  The parameter
$NU_{11}=17.2$, and the projection onto the exact stationary state at
this value varies between 60\% and 95\%.  The efficiency, i.e., the
ratio of final and initial atom number, is 9.5\%. The comparison with
the eigenstate of the linear case (see inset) shows a broadening of
the wave function and a significant reduction of the central density
due to the interatomic repulsion.

\subsection{Collapse, revival, and free expansion}

\begin{figure}
\centerline{\psfig{figure=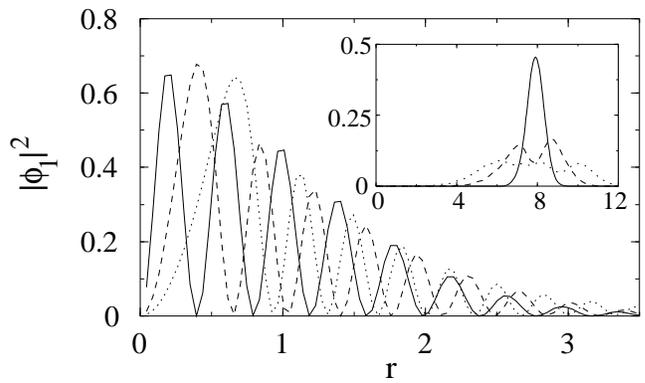,width=8.6cm,clip=}}
\caption{Condensate density $|\phi_1|^2$ at times $\Delta t=1.6$
and 3.2 (inset) after suddenly switching off the RF coupling $\Omega$. 
The different curves have nonlinearity parameters $UN=0$ (full curve),
5 (dashed), and 20 (dotted). In each case, the initial state was the
respective matter-wave bubble at $\Delta=60$ and $\Omega=9$.}
\label{figcoll}
\end{figure}

Another interesting effect occurs if, after creating the bubble, the
coupling strength $\Omega$ is instantaneously reduced to zero. In this
case, the two components of the bubble evolve almost independently of
each other in their respective bare potentials.  Component 2 is
therefore rapidly expelled from the trapping region, whereas component
1 undergoes a collapse or contraction into the center of its binding
harmonic potential followed by a re-expansion. If non-linear
interactions can be neglected this scenario repeats itself
periodically, the wave function regaining its initial shape at times
$n\pi$, $n=1,2,\dots$ In the presence of atomic interactions, however,
the shape is gradually distorted. Figure \ref{figcoll} shows
$|\phi_1(r)|^2$ for three different values of $gN$ at times $t=1.6$
and 3.2, i.e., for `complete' collapse and re-expansion, respectively.
In the large figure the wave functions show an interference pattern
which is due to particles from opposite sides of the bubble passing
through each other. Note that for growing non-linear interaction the
central interference fringes are pushed outwards in agreement with
earlier studies on similar systems (see, e.g., \cite{LiuWuNu00}).  It
might be possible to infer the nonlinearity parameter from the fringe
pattern. The inset in Fig.\ \ref{figcoll} illustrates the subsequent
broadening of the wave function in the course of the re-expansion due
to non-linear interactions.

\begin{figure}
\centerline{\psfig{figure=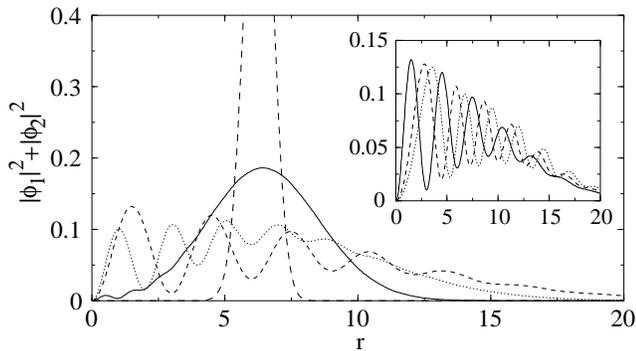,width=8.6cm,clip=}}
\caption{Condensate density $|\phi_1|^2+|\phi_2|^2$ at times
$\Delta t=0.0$ (long-dashed), 2.0 (full), 4.0(dotted), and 6.0 
(short-dashed)
after suddenly switching off both the RF coupling and the magnetic
potential. The initial state is the matter-wave bubble
at $\Delta=40$, $\Omega=9$, and $UN=0$. In the inset the condensate
density is shown at $\Delta t=6.0$ for nonlinearity parameters $UN=0$
(full curve), 20 (dashed), and 40 (dotted).}
\label{figexp}
\end{figure}

A related behavior can be observed if the magnetic fields are switched
off along with the RF coupling. As it is no longer subject to any
potential, the localized radial bubble wave function displays a time
evolution similar to a free particle and gradually broadens due to
dispersion (see Fig.\ \ref{figexp}). After some time wave function
pieces from opposite sides of the bubble start to overlap each other
and an interference structure ensues. Again, the interference fringes
are shifted outwards if non-linear interactions become significant
(see the inset).

\section{Disc-shaped condensates}
\label{sec:discs}

In this section we examine how our trapping scheme is modified in the
presence of gravity and in the absence of any gravitational
compensation (such as described in section \ref{subsec:bubbgrav}).
The results will also lay the foundation for our proposal for the
creation of 2D atom traps which is outlined in the following section
\ref{sec:2dtrapping}.

\subsection{Adiabatic trapping in the presence of gravity}

To make the presentation more concrete, we look at a specific example
that corresponds to typical experimental conditions.  We thus consider
a condensate of about 10$^5$ Rb87 atoms initially prepared in the
$F=2$, $M_F=2$ hyperfine sublevel and in the ground state of an
anisotropic magnetic trap with frequencies $\nu_y=\nu_z=220$ Hz and
$\nu_x=11$ Hz. The $z$ axis again points along the vertical direction.
An RF field is then applied that couples the sublevels within the
$F=2$ multiplet. By appropriately tailoring the time dependence of the
field, the BEC always remains in the ground state of the RF-induced
adiabatic potential and is thus manipulated in a controlled way.

The condensate dynamics is determined by the Gross-Pitaevskii equation
for the components $\psi_M$, $M=-2,\dots,2$,
\begin{eqnarray}\label{GPE1}
i\dot\psi_M&=&\left[-\frac {\bm{\nabla}^2} 2 +\frac
M4(\kappa^2 x^2+
y^2+z^2)-M\Delta(t)+Gz\right]\psi_M \nonumber \\
 & & +\Omega_{M,M-1}(t)\psi_{M-1}
 +\Omega_{M+1,M}(t)\psi_{M+1} \nonumber \\
 & & +UN\rho(r,t)\psi_M.
\end{eqnarray}
Again, all quantities are dimensionless.
They are now scaled to units derived from the
radial angular frequency $\omega_z=2\pi\nu_z$.
The effective detuning is given by $\Delta(t)=[\hbar\omega_{f}-\Delta 
E(0)]/
\hbar\omega_z$ with
$\Delta E(0)$ the Zeeman energy split between subsequent
hyperfine sublevels at the magnetic field minimum.
The parameter $\kappa=\nu_x/\nu_z$ gives the ratio between axial and
radial trapping frequencies. The scaled gravitational
acceleration $G$ becomes 7.06 in the present setup.
The RF coupling constants are given by
$\Omega_{2,1}=\Omega_{-1,-2}=\sqrt{2/3}\Omega_{1,0}=
\sqrt{2/3}\Omega_{0,-1}=\Omega(t)$. For simplicity, all non-linear
interactions coefficients are taken to be equal and denoted by $U$;
$N$ is the total number of atoms and $\rho(r,t)=\sum_{M=-2}^2|\psi_M
(r,t)|^2$.

As discussed in Sec.\ III.C, the simplified form of the non-linear
interaction allows us to transform to
the dressed-eigenstate basis in the same way as in the linear case.
The field-induced adiabatic potentials are now given by
\begin{equation}\label{VdressedM}
 \tilde V_{M}({\bf r},t)=M\sqrt{[V_{tr}({\bf
 r})-\Delta(t)]^2+\Omega^2(t)},
 \end{equation}
where $V_{tr}({\bf r})=(\kappa^2 x^2+y^2+z^2)/4$,
so that a wave packet initially prepared in the state $M=2$ can  
evolve in the potential $\tilde V_{2}({\bf r},t)+Gz$
if the potential is deformed slowly enough. The approximate equation of
motion  for ${\tilde\psi}_2$  thus reads
 \begin{eqnarray}\label{GPE2}
 i\dot{\tilde\psi}_2&=&\left[-\frac {\bm{\nabla}^2} 2
 +\tilde
 V_{2}({\bf r},t)+Gz+UN|{\tilde \psi}_2({\bf r},t)|^2\right]{\tilde
 \psi}_2.
 \end{eqnarray}
 
\begin{figure}
\centerline{\psfig{figure=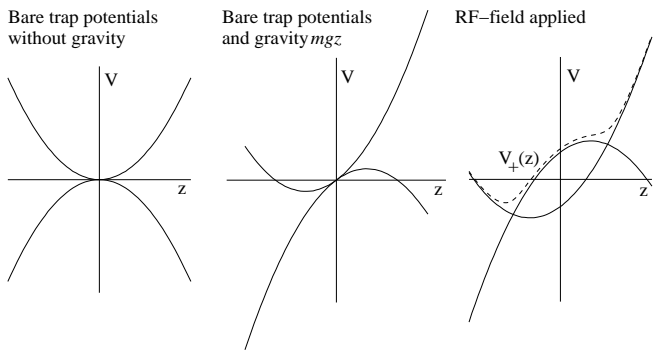,width=8.6cm,clip=}}
\caption[f1]{Schematic of the potentials along the $z$-direction
showing the characteristic influences of gravity and the applied
RF-field.}
\label{gravpot}
\end{figure}

 The influence of gravity becomes apparent by writing the z-dependent
 part of the potential as $Mz^2/4+Gz= M(z+2G/M)^2/4-G^2/M$. We see
 that, effectively, the position of the bare trap minimum is shifted
 to $z=-2G/M$ (which typically is large compared to the ground state
 extension), the minimum itself is shifted by $-G^2/M$. At $\Delta=0$
 all bare potentials touch at $z=0$. The bare potential for $M=-2$
 intersects the $M=2$ potential at its minimum for $\Delta=G^2/4$. As
 the condensate is initially localized at this minimum, we conclude
 that only for $\Delta\agt G^2/4$ does a significant shift and
 deformation of the ground state set in.  For large $\Delta$ the
 ground state is located at about $2\sqrt{\Delta}$.  In Fig.\ 
 \ref{gravpot} we show cuts along the potentials in the $z$ direction
 that illustrate the characteristic behavior explained above.
 
 What does the full spatial dependence of the potential $\tilde V_{2}({\bf
   r},t)+Gz$ look like? The potential $\tilde V_{2}({\bf r},t)$ alone has
 its minimum on an ellipsoidal surface defined by $V_{tr}({\bf
   r})=\Delta(t)$.  However, due to the influence of gravity the combined
 potential $\tilde V_{2}({\bf r},t)+Gz$ is strongly tilted in the $z$
 direction, so that the wave function will assemble around its bottom (we
 will see this in Fig.\ \ref{fig3a}).  The main control parameter to vary
 the shape of the adiabatic potential is the detuning $\Delta$. At a given
 $\Delta$ the coupling strength $\Omega$ has to be chosen large enough that
 the lifetime of the condensate becomes sufficiently long.

\subsection{Preparation}
\label{subsec:discprep}

\begin{figure}
\centerline{\psfig{figure=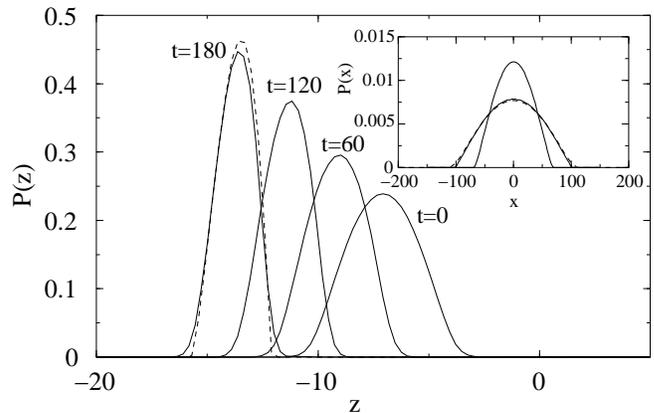,width=8.6cm,clip=}}
\caption[f1]{Loading an adiabatic trap in the presence of gravity. Shown
is the $x$-integrated density $P(z)=\int dx\sum_{M=-2}^2|\psi_M(x,z)|^2$
at the indicated times for the preparation process described in the text. 
The
inset shows the $z$-integrated density $P(x)$ at $t=0$ and 180. The
dashed curves show $P$ as obtained from the Thomas-Fermi approximation.}
\label{figprep}
\end{figure}

To demonstrate that atoms can be actually be trapped in the RF-induced
adiabatic potential we have performed numerical simulations of Eq.\ 
(\ref{GPE1}) in two dimensions, i.e., the direction of gravity $z$ and the
weak trapping direction $x$. These calculations should be able to capture
the main aspects of the wave packet behavior. Our results indicate that the
same two-step approach as outlined in Sec.\ III can be used to transfer the
BEC into the adiabatic trap. As an illustration, we show in Fig.\ 
\ref{figprep} the result of one of our simulations. For the 2D calculation,
the nonlinearity parameter $NU$ was chosen so that the extension of the
ground state in the bare magnetic trap [which is the initial state
$\psi_2(t=0)$] coincides with the extension in $x$ and $z$ of the 3D ground
state for the given atom number of 10$^5$.  In the first step of the
preparation scheme, the RF intensity is linearly ramped up to the desired
final value of $\Omega$ at {\it negative} $\Delta$.  In Fig.\ \ref{figprep},
in order to reach a final $\Omega=12$ ($\approx 2.64$kHz) the field is
switched on at $\Delta=-5(\approx -1.1$kHz) within $\Delta t=20\simeq
14.4$ms.  In the second step the RF detuning is simply increased to the
final value thereby keeping the intensity fixed. In Fig.\ \ref{figprep}, the
detuning $\Delta$ was increased to $60(\approx 13.2 $kHz) within $\Delta
t=160 \simeq 116$ms.  The simulations already show the deformation effect
that we build upon in the next section to obtain 2D trapping, namely, a
squeezing in the $z$ direction indicating tighter trapping, and an expansion
in $x$ showing the decrease in the corresponding trap frequency.
Furthermore, we see that the Thomas-Fermi approximation gives a very good
description of the wave function. Nevertheless, it should be mentioned that
during preparation a slight excitation of the wave function along the $x$
axis appears (not visible in Fig.\ \ref{figprep}) indicating some
nonadiabatic effects.

The effort required for the numerical calculations shows that it is very
difficult to go significantly beyond the trap deformations shown in the
above example---and in particular to simulate reaching the 2D limit.  In the
next section we give some semi-quantitative arguments to determine the
values of $\Omega$ and $\Delta t$ that should be used to reach a given
$\Delta$. Here we restrict ourselves to a qualitative discussion.  The
chosen $\Delta$ places a lower limit on the value of $\Omega$---only for a
high enough $\Omega$ is the life-time of the BEC in the adiabatic potential
sufficiently long. For this complicated five-state system it is difficult to
estimate the lifetimes analytically in an accurate way. However, the
numerical calculations indicate that with, for example, $\Delta=60$, a value
of $\Omega=12$ should be adequate.  Furthermore, the study of the linear
two-state system shows that the lifetime increases exponentially with
$\Omega$. We thus expect that the lifetime can always be adjusted by a
moderate increase in $\Omega$.

Concerning the rise times $\Delta t$, they are bounded from below by the
requirement that we want to avoid nonadiabatic excitations and keep the BEC
in the ground state. We expect that it is more likely we would generate such
excitations along the $x$ direction than along the radial directions as in
the former the eigenstates are spaced much more closely. In fact, this is
observed in the numerical simulations as mentioned above.  The amount of
excitation is reduced by performing the process more slowly. An upper limit
on the rise time is placed by the lifetime of the adiabatic ground state and
the overall decay time of the BEC.

\subsection{Signatures of adiabatically trapped BECs}

There are various signatures which allow one to verify experimentally that
the condensate behavior is indeed determined by the adiabatic potential
(apart from the fact that the BEC is still there at all):

1.\ The vertical position of the condensate is changed (lowered).
With $z=0$ indicating the minimum of the magnetic trap field the
condensate is located around $z=-G$ ($\approx -5.1\mu$m for our
example) in the absence of the applied RF fields. With these fields
turned on at sufficiently large $\Delta$ the potential minimum is
shifted to the vicinity of the intersection between the different bare
potentials at $x=y=0$, $z=-2\sqrt{\Delta}$. For $\Delta=60$, e.g., the
condensate center is transferred to $-11.3\mu$m, i.e., it is shifted
over a distance larger than the BEC extension in the $z$ direction.

2.\ The shape of the condensate is changed. At $\Delta=40$,
$\Omega=12$ the full width $\sigma_z$ in $z$ is about half the width
of the RF-field free case\footnote{A rough estimate for $\sigma_z$ can
be obtained by expanding the adiabatic potential around its minimum
and then applying the Thomas-Fermi approximation. Thereby, one
obtains $\sigma_z\approx \sqrt{\tilde\mu\Omega/\Delta}$ with
$\tilde\mu$ the new chemical potential (with respect to the
potential minimum), which is typically somewhat higher than the
original $\mu_0$ ($\simeq 6.17 \hbar\omega_z$ in the present
case).}, whereas the width in $x$ has increased by about 50\%.  This
change of shape should influence the ballistic expansion of the BEC
once the trapping fields are switched off.  Furthermore, the
condensate is slightly bent in the $z$ direction.

3.\ The ground state in the adiabatic potential is a dressed state,
i.e., it is a superposition of different hyperfine sublevels.
Observing the different hyperfine components would very
convincingly demonstrate the trapping of the BEC in the adiabatic
potential.

4.\ If the RF fields are suddenly switched off the components $M_F=1$
and 2 should perform harmonic oscillations in their respective magnetic
trapping potentials.

\section{Two-dimensional atom trapping regime}
\label{sec:2dtrapping}

\subsection{Basic considerations for 2D atom traps and 2D BEC}
\label{subsec:2dtraps}

Having introduced field-induced adiabatic potentials as a means to
create new types of trapping potentials, we now turn to the question
of how they can be used to obtain a two-dimensional atom trap by the
appropriate choice of strong fields and large detunings.
Unfortunately, as mentioned above, in these parameter regimes the
resulting systems are hard to model numerically.  So in this section
we will use general arguments to show under what conditions it might
be possible to obtain 2D trapping, and even a 2D BEC, using the kind
of adiabatic loading scheme mentioned in Sec.\ \ref{sec:discs}.  These
arguments will allow us to relax the restriction on the geometry of
the system somewhat (the magnetic trap can now have three different
frequencies). However, in our discussion we explicitly consider a
two-state system (rather than general multi-level systems).
Nevertheless we expect that much of the discussion of a two-state
system would hold qualitatively for multi-level systems with an
appropriate choice of parameters.

In general,
a 3D harmonic
potential with angular frequencies $\omega_1$, $\omega_2$,
$\omega_{trans}$
provides an effective two-dimensional trap for atoms of temperature $T$, if
\begin{equation}
\hbar \omega_{1,2} < k_B T < \hbar \omega_{trans}
\label{eq:2Dcondition}
\end{equation}
(from now on we return to
unscaled quantities). Thus a good 2D trap has a large $\omega_{trans}$ to
allow one motional degree of freedom to be frozen out at high
temperatures. For an ideal gas, Bose condensation in a 2D harmonic trap
occurs at
\begin{equation}
  k_B T_c =  \hbar \bar\omega \sqrt{6 N} / \pi
\label{eq:2Dtemp}
\end{equation}
with $\bar\omega=(\omega_1\omega_2)^{1/2}$ and $N$ the number of atoms
\cite{BagKle91}.  Thus, combining Eq.\ (\ref{eq:2Dcondition}) and Eq.\ 
(\ref{eq:2Dtemp}), the number of atoms that can undergo a genuine 2D
condensation satisfies the condition $N\alt
(\omega_{trans}/\bar\omega)^2$.  For a higher number of atoms
condensation would have occurred already in the 3D regime.  Therefore
we see that, to aim for a 2D BEC, a high ratio
$\omega_{trans}/\bar\omega$ is a further desirable criterion for a 2D
trap, especially considering that small numbers of atoms would be hard
to image experimentally.

\subsection{Two-dimensional atom trapping and adiabatic potentials}
\label{subsec:2dadia}

We will generalize the discussion in Sec.\ \ref{sec:discs} so that we
now allow the original magnetic trap to have different oscillator
frequencies $\omega_x$, $\omega_y$, and $\omega_z$ along the three
axes with corresponding oscillator lengths $a_x$, $a_y$, $a_z$. (Note
that $a_i = \sqrt{\hbar/m\omega_i}$, $i=x,y,z$).  Thus the
corresponding magnetic trap has the potential
\begin{equation}
  V_{tr}({\bf r})= \frac{1}{2} m  (\omega_x^2 x^2 +\omega_y^2 y^2
  +\omega_z^2 z^2)
 \label{2state:Vtr}
\end{equation}
where, again, we are now working with unscaled quantities. As in Sec.\ III, 
we consider a two-level system and assume that the second state experiences
the magnetic potential $-V_{tr}$.

The upper field-induced adiabatic potential, which is the potential
of greatest interest, is then given by
\begin{equation}
 \tilde V({\bf r},t)= mgz +  \tilde V_{+}({\bf r},t)
 \label{2state:Vall}
 \end{equation}
with
\begin{equation}
 \tilde V_{+}({\bf r},t)=  \sqrt{[V_{tr}({\bf
 r})-\hbar\bar\Delta(t)/2]^2+ (\hbar\Omega(t))^2 } .
 \label{2state:Vdressed}
 \end{equation} 
 Here $\bar\Delta$ denotes the unscaled detuning, i.e.\ 
 $\bar\Delta=\omega_f -\Delta E(0)/\hbar$ with $\omega_f$ the RF field
 frequency and $\Delta E(0)$ the minimum energy difference between the
 hyperfine states (see Sec.\ II.A).  The `bare' potentials intersect,
 where there is resonance, i.e., at the locations $\hbar\bar\Delta(t)
 = 2 V_{tr}({\bf r})$. This intersection we call the `seam' of the
 bare potentials.  In the absence of gravity, the seam of $\tilde V_+$
 would also be the location of the minimum in $\tilde V$ and would
 have the shape of an ellipsoid with radii
 $r_i=(\bar\Delta/\omega_i)^{1/2}a_i$, $i=x,y,z$.  Thus, without
 gravity, or gravity being compensated for, the atom distribution
 forms an ellipsoid bubble, or shell in the potential
 (\ref{2state:Vdressed}).
 
\begin{figure}
\centerline{\psfig{figure=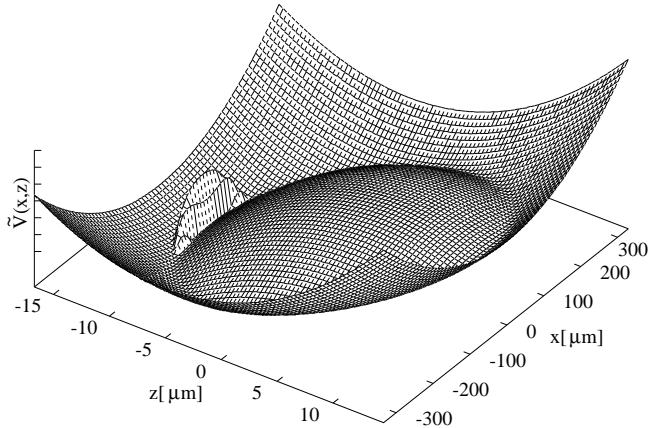,width=8.6cm,clip=}}
\caption[f1]{Adiabatic potential $\tilde V({\bf r})=\tilde V_+({\bf r})+mgz$ (in
arbitrary units) at $y=0$ for $\omega_x/2\pi=11$ Hz,
$\omega_y/2\pi=\omega_z/2\pi=
220$ Hz, $\bar \Omega/2\pi=2.6$ kHz, and $\bar \Delta/2\pi=66$ kHz. Inserted is the
ground state
of a Rb87 condensate with about 10$^5$ atoms.
}
\label{fig3a}
\end{figure}

 Under the influence of gravity the atoms sag to the bottom of the shell
 potential (\ref{2state:Vall}), and this would not allow the formation of a
 matter-wave bubble.  In fact, as shown in Sec.\ \ref{subsec:bubbgrav}, a
 closed bubble could exist only up to radii $r\alt(\hbar^2/gm^2)^{1/3}$
 ($\simeq 5\times 10^{-7}$m for Rb87). Under typical experimental conditions
 we can expect a situation as depicted in Fig.\ \ref{fig3a} which
 shows---for the indicated parameter values---the potential $\tilde V$ in
 the plane $y=0$ along with an atomic BEC in the ground state for this
 potential.

Typically, an atom moving in the potential $\tilde V$ is confined to the
close proximity of the seam due to the strong confinement in the direction
transverse to it. If its energy energy $E$ satisfies $E\ll mgr_z$, it is
also restricted to the vicinity of the bottom of $\tilde V$ because of
gravity. To a good degree of approximation the atomic motion can therefore
be modelled as harmonic. To determine the harmonic frequencies we can expand
Eq.~(\ref{2state:Vall}) about $z=-r_z$ to find
\begin{equation}
  \omega_{1,2} = (g/r_z)^{1/2}\omega_{x,y}/\omega_z
\label{2d:om12}
\end{equation}
along the surface of the seam, and
\begin{equation}
  \omega_{trans}=(\bar\Delta/\bar\Omega)^{1/2}\omega_z
\label{2d:omtrans}
\end{equation}
in the normal direction.
Here the unscaled coupling frequency is denoted by  $\bar\Omega$.
The oscillator lengths corresponding to $\omega_{1,2}$ and $\omega_{trans}$
will be denoted by $a_{trans}$, $a_1$, and $a_2$. The effect of gravity on
the potential will also shift it downwards so that the equilibrium point
is now at
\begin{equation}
 z_0 \sim - \sqrt{\frac{ \hbar \bar\Delta }{ m \omega_z^2 }}
 - \frac{ g }{  \omega_{trans}^2 } .
  \label{eq:2dshift}
\end{equation}
However, the displacement of $g/ \omega_{trans}^2$ is rather small and also
turns out not to be sufficient to violate the harmonic expansion of
Eq.~(\ref{2state:Vall}) which results in Eqs.~(\ref{2d:om12}) and
(\ref{2d:omtrans}).
(See section \ref{subsub:harmonic} below.)

\subsection{Conditions for 2D trapping in adiabatic potentials}

The obvious strategy for realizing a 2D trap is to increase the
detuning $\bar\Delta$ as much as possible in order to obtain strong radial
confinement. A sufficiently large coupling $\bar\Omega$ will be necessary to
ensure that the potentials remain adiabatic.  To arrive at a 2D trap, a
number of considerations have to be taken into account (some given
previously in Ref.\ \cite{ZobGar00a}), and these factors determine how
large $\bar\Omega$ and $\bar\Delta$ need to be, and the relationship between
them.

\subsubsection{Lifetime}
\label{subsub:lifetime}

For a given $\bar\Delta$, the RF coupling strength $\bar\Omega$ has to be
chosen accordingly to assure a sufficiently large lifetime for
atoms held in the adiabatic trap.
Equation (\ref{gamma2}) gives an estimate of the decay rate
and if we require that the decay rate is less than a certain maximum,
i.e.,
\begin{equation}
  \label{eq:gamma}
  \gamma_{ex} \le \gamma_{max}
\end{equation}
it follows from Eq.~(\ref{gamma2}), when unscaled, that one needs
\begin{equation}
\bar\Omega^3\ge \lambda \omega_z^2\bar\Delta
 \label{cond:lifetime} 
\end{equation}
with the parameter $\lambda=2[\ln(\gamma_{max}/2\omega_z)]^2/\pi^2$.
Thus the worst allowed decay rate can be specified in terms of
the vertical trap frequency.
In the numerical example below (Sec.\ \ref{sec:2dnumerical})
we choose $\gamma_{max}/\omega_z=0.01$ so that $\lambda=5.7$.

As indicated in Sec.\ V.B, analytical expressions for the lifetime in
general multi-state systems are not available. However, we expect that our
estimates are still valid, at least as a first approximation, under such
conditions. This view is supported by our numerical study of the five-state
system.

\subsubsection{Strong binding and spatial thinness}
\label{subsub:strongthin}

The lifetime sets a lower limit on $\Omega$ at a given $\Delta$. Choosing
$\Omega$ too large, however, impairs the strong radial confinement.
An upper limit is imposed by the obvious conditions
\begin{equation}
  \omega_{trans} \gg \omega_{1,2}
\label{eq:banding}
\end{equation}
and
\begin{equation}
   r_z\gg a_{trans} .
\label{eq:thin}
\end{equation}
The first of these was discussed in Sec.\ \ref{subsec:2dtraps} and
ensures that one spatial direction is effectively frozen out at low temperatures.
The second condition guarantees that the atoms are tightly confined to the vicinity
of the seam.
Both these requirements are typically not in conflict with condition
(\ref{cond:lifetime}). Formally, this can be seen from
the fact that the first condition translates into
\begin{equation}
   \bar\Omega^2 \ll \frac{ \bar\Delta^{3}   \omega_z^{5} a_z^2 }{
                           g^2 \omega_{x,y}^4 },
\label{cond:banding} % B13
\end{equation}
where for $\omega_{x,y}$ one should choose the maximum of $\omega_{x}$ and
$\omega_{y}$ to make the condition most restrictive.
The second condition leads to
\begin{equation}
   \bar\Omega \ll 4 \bar\Delta^{3} / \omega_z^2 .
\label{cond:thin} %B9
\end{equation}
Inequality (\ref{cond:lifetime}) will allow us to choose $\bar\Omega \propto
\bar\Delta^{1/3}$, so that the above conditions are increasingly easy to
meet for growing $\bar\Delta$, in accordance with our expectation.

%%%%%%%%%%%%%%%%%%%%%%%%%%%%%%%%%%%%%%%%%%%%%%%%%%%%%%%%%%%%%%%%%%%%%%

\subsubsection{Harmonic 2D trapping}
\label{subsub:harmonic}

In this section we will look at the conditions for harmonicity
of the trap. Harmonicity is probably desirable, but is not actually
\textit{essential} for trapping. The conditions derived below will show
that the trap is operated in a 2D harmonic regime.

To be able to speak of a {\it harmonic} 2D trap at a finite temperature, 
one
has to require that 
\begin{equation}
  \hbar\omega_{1,2} \ll k_B T_c \ll mgr_z .
\label{eq:require:harmonic}
\end{equation} 
The first
criterion ensures that, at the condensation point, the atoms can reach
sufficiently many quantum states to actually experience the trapping
potential as being harmonic. It is automatically fulfilled if we substitute
for $T_c$ from Eq.~(\ref{eq:2Dtemp}) to find
\begin{equation}
    N \gg  \frac{\pi^2}{6}\left( \frac{\omega_{1,2}}{\omega_{2,1}} \right) ,
  \label{eq:B11}  
\end{equation}
where from Eq.~(\ref{2d:om12}) 
$\omega_{1,2}/\omega_{2,1} = \omega_{x,y}/\omega_{y,x}$ is the ratio of
trapping frequencies in the $x$-$y$ plane. Equation (\ref{eq:B11}) should
be easily satisfied if there are to be any significant number of atoms in
the trap. 

The second part of Eq.~(\ref{eq:require:harmonic}) leads to
\begin{equation}
  k_B T_c \ll \frac{ g \bar\Delta^{1/2} }{  a_z \omega_z^{5/2}  }
  \label{eq:flatness}  %% B10
\end{equation}
when we substitute for $r_z$.  This second criterion makes sure that
anharmonic effects in the trapping potential can be neglected.  The same
condition arises as the condition for harmonic motion of a single particle
confined
to the `seam'. (A picture which assumes tight transverse trapping.)  In the
example below $mgr_z/k_B \simeq 18$ $\mu$K which is much larger than
$T_c\simeq 0.3$ $\mu$K. Note, however, that the appearance of anharmonic
effects in the 2D trapping potential does not automatically affect the
two-dimensionality of the trap.

We can also ask if we have harmonicity in the third,
transverse, direction. If we focus on the bottom of the potential, i.e.\
take $x=y=0$, the adiabatic potential (\ref{2state:Vall}) simplifies
to
\begin{equation}
 \tilde V(z,t)= mgz +  \sqrt{\left(  m \omega_z^2 z^2 /2
  -\hbar\bar\Delta(t)/2 \right)^2+ (\hbar\Omega(t))^2 } .
 \label{2state:Vall_xy0a}
 \end{equation}
Then in an adiabatic regime, one limit which produces
quite simply a harmonic potential is found from the condition
\begin{equation}
 \hbar\Omega(t)  \gg \left| \frac{1}{2}  m \omega_z^2 z^2  
  -\hbar\bar\Delta(t)/2 \right| ,
\label{eq:harm_trans_raw}
 \end{equation}
which together with
\begin{equation}
 |z-z_0| /|z_0| \ll 1 ,
 \label{eq:dzcond}
 \end{equation}
allows a quadratic expansion of Eq.~(\ref{2state:Vall_xy0a}).
An estimate for $z_0$, the location of the minimum, has
been given in Eq.~(\ref{eq:2dshift}).
At $z=z_0$ we may
substitute Eq.~(\ref{eq:2dshift}) into Eq.~(\ref{eq:harm_trans_raw})  
to eventually obtain
\begin{equation}
  \bar\Delta \gg    \frac{ g^2 }{  a_z^2 \omega_z^{3}  } .
  \label{eq:harm_exp_ok} %% B7A
\end{equation}
Here we also assumed that the shift of $z_0$ away from
$ -r_z=- \sqrt{ \hbar \bar\Delta / m \omega_z^2 }$ (see Eq.~(\ref{eq:2dshift}))
was very small so that the smallest term could be dropped from $z_0^2$.
The condition for this approximation is that
\begin{equation}
 \sqrt{\frac{ \hbar \bar\Delta }{ m \omega_z^2 }} \gg
   \frac{ g }{  \omega_{trans}^2 }
 \label{eq:smallshiftpre} 
\end{equation}
which would lead to the subsidary condition
\begin{equation}
  \bar\Delta^3 \gg    \frac{ g^2 \bar\Omega^2 }{  a_z^2 \omega_z^{3}  } ,
  \label{eq:smallshift} %%B7B
\end{equation}
when we substitute for $\omega_{trans}$. 

In considering values of $z$ away from $z_0$ in Eq.~(\ref{eq:dzcond}) and
Eq.~(\ref{eq:harm_trans_raw}), we need an estimate of the thickness of the
matter-wave disc. A distance characterising this could be $a_{trans}$, i.e.\ 
we would use a single particle wave function width as a measure of the size
of $|z-z_0|$ in Eq.~(\ref{eq:dzcond}). This equation is then already
satisifed by Eq.~(\ref{eq:thin}) and the remaining condition
Eq.~(\ref{eq:harm_trans_raw}) is also, approximately, satisifed if we accept
the scaling of Eq.~(\ref{cond:lifetime}),
(with a typical value of $\lambda$).

It turns out that it is possible to satisfy all the constraints in this
section. It is even the case that violation of Eqs.~(\ref{eq:B11}) and
(\ref{eq:flatness}) need not prevent a 2D BEC since there is no reason why a
BEC cannot be 2D in an anharmonic regime; equation (\ref{eq:2Dtemp}) would
simply not apply. Likewise, anharmonicity in the transverse direction,
resulting in violation of conditions (\ref{eq:harm_exp_ok}) and
(\ref{eq:smallshift}) need not prevent 2D trapping or a 2D BEC.

\subsubsection{Loading}
\label{subsub:loading}

A straightforward way to load the 2D trap consists in starting from a
condensate in the original magnetic potential and then adiabatically
transferring it to the 2D trap by appropriately switching on the RF
field.
The minimum duration of this process can be estimated by stipulating the
adiabaticity condition $\dot\omega_{1,2} \ll \omega_{1,2}^2$. Using
relation (\ref{cond:lifetime}) this leads to
\begin{equation}\label{time}
t\gg \frac 1 {\omega_{1,2}} = 
 \frac{(\bar\Omega/\omega_{x,y})^{3/4}}{\lambda^{1/4}\sqrt{g/a_{x,y}}}.
\end{equation}
Note that during loading, $\bar\Omega$ can be increased above the intended
final value to reduce intermediate adiabatic losses.

\subsubsection{Temperature}  
\label{subsub:temperature}

For the trap to be 2D at a temperature $T$ the second part of
Eq.~(\ref{eq:2Dcondition}) applies ($k_B T < \hbar \omega_{trans}$)
which with the substitution of Eq.~(\ref{2d:omtrans}) leads to
\begin{equation}
   k_B T \ll \sqrt{\frac{ \bar\Delta}{ \bar\Omega} } \hbar\omega_z
  \label{eq:transverse_spacing}
\end{equation}
If we take the case of equality in Eq.~(\ref{cond:lifetime})
and substitute for $\bar\Delta$ we find
\begin{equation}
    k_B T \ll   \hbar\bar\Omega/\sqrt{\lambda}.
  \label{eq:transverse_spacing_2} %% B12 %%
\end{equation}
We would expect this condition to be more restrictive than
Eq.~(\ref{eq:flatness}) which leads to 
$  k_B T \ll  \hbar\bar\Omega^{3/2} g / 2\sqrt{\lambda} a_z  
\omega_z^{7/2}$.
Thus, by substituting Eq.~(\ref{eq:2Dtemp}) into 
(\ref{eq:transverse_spacing_2}),
the number of atoms that can undergo a 2D condensation is limited by
\begin{equation}
  N \ll \frac{ \pi^2 \bar\Omega^2 }{ 6 \lambda \bar\omega^2 } .
\end{equation}
If we now utilize $\bar\omega=(\omega_1\omega_2)^{1/2}$ and
Eq.~(\ref{2d:om12}) we finally obtain
\begin{equation}
N  \ll \frac{\pi^2 a_z \bar\Omega^{7/2}\omega_z^{1/2}}
{6g\lambda^{3/2}\omega_x\omega_y}.
\label{eq:final_N_constraint}
\end{equation}
Since Eq.~(\ref{eq:final_N_constraint}) puts an upper bound on the number of
atoms it is desirable to have as large a coupling $\bar\Omega$ as possible.

In fact none of the conditions above constrain $\bar\Omega$ if the
connection (\ref{cond:lifetime}) is accepted. The higher the value of
$\bar\Omega$ the better (as in Eq.~(\ref{eq:final_N_constraint})).  However,
from a practical point-of-view, it would be desirable to find the lowest
viable $\bar\Omega$. This is determined by the most restrictive inequality
on $\bar\Omega$. That is, if we place each of our various inequalities in
the form $\bar\Omega > k$, where $k$ contains the remaining constants, we
will look for the inequality with the largest constant $k$.  For realistic
values of $\lambda$ and $\omega_{x,y}$ this appears to be
Eq.~(\ref{eq:harm_exp_ok}), which on substitution of
Eq.~(\ref{cond:lifetime}) leads to
\begin{equation}
   \bar\Omega^7 \gg  \lambda^3 g^2 \omega_z^3 / a_z^2 
    .
\label{eq:omega_final}
\end{equation}
Any value of $\bar\Omega$ greater than this would be suitable, but note that
increasing $\bar\Omega$ also increases the loading time, Eq.~(\ref{time}),
and this is ultimately undesirable.  To see what is realistically possible
we must now determine some values for a practical case.

\subsection{Numerical estimate}
\label{sec:2dnumerical}

We reported a numerical estimate in Ref.\ \cite{ZobGar00a}, where we
considered a typical Ioffe-Pritchard trap with $\omega_x/2\pi=11$ Hz,
$\omega_y/2\pi=\omega_z/2\pi=220$ Hz containing Rb87 atoms initially in the
$F=2$, $M=2$ ground state.
In order to match approximately the two-state theory given above to
this five-state system we need to replace $\bar\Delta$ by $4 \bar\Delta$
to obtain the correct condition for resonance. The trap potential
(\ref{2state:Vtr}) becomes that for the $M_F=2$ state, and the coupling
$\bar\Omega$ is replaced by $2\bar\Omega$.
Then an RF field with a final coupling $\bar\Omega/2\pi=15$ kHz and [from
(\ref{cond:lifetime})] $\bar\Delta/2\pi=12.2$ MHz is one that 
can be provided with currently
available technology \cite{Bos00}. 
The condition~(\ref{eq:omega_final}) is easily satisfied
since it results in $\bar\Omega/2\pi \gg 405$ Hz.
The resulting trap frequencies are
$\omega_{trans}/2\pi=8.9$ kHz, $\omega_1/2\pi=1.3$ Hz, $\omega_2/2\pi=27$ 
Hz. The new
trap is vertically shifted by 0.34 mm from the center of the original
magnetic potential. The critical temperature is given by 0.43$\mu$K, so
that up to 3.6$\times 10^6$ atoms could be condensed.
The transverse width of the condensate is estimated at 0.08$\mu$m
if atomic interactions can be neglected. The time required for the
preparation process should be large compared to $1/\omega_1$,
Eq.~(\ref{time}), i.e., of
the order of several seconds. In view of these estimates,
the experimental realization of this new kind of 2D trap seems
to be within reach of current experiments.

\section{Summary and conclusions}

In this paper we have shown that field-induced adiabatic potentials can
provide a robust and versatile tool to create novel trapping configurations
for ultracold atoms. As specific examples, we have considered the generation
of matter-wave shell, or bubble, states and the preparation of
two-dimensional atom traps. In a bubble state, the matter-wave density is
localized around the surface of a three-dimensional sphere.  We have
discussed the preparation process which proceeds by coupling a harmonic
magnetic trap to a repulsive harmonic potential by a suitably chosen
time-dependent RF field. Although this configuration appears to be unstable
at first sight, the bubbles are stabilized in the resulting dressed
potential and their lifetime increases exponentially with the RF coupling
strength.  The bubbles can be used as stepping-stones for the creation of
highly-excited oscillator eigenstates and `non-linear eigenmodes.' We also
investigated possible experiments showing collapse and revival effects.
Although the creation of the bubbles is impeded by the influence of gravity
and trap anisotropies, we have pointed out ways to overcome these
difficulties with the help of present-day technology.

The same principle, which has been used for making bubble states, can also
be applied to the creation of two-dimensional atom traps. Under the combined
influence of the dressing RF field and gravity, a condensate pools at the
bottom of the resulting potential. By increasing the RF detuning the radial
confinement becomes steeper, and the condensate gets increasingly squeezed
until it eventually reaches a `quasi-two-dimensional' state. We have given
estimates for the parameters necessary to reach this regime.

We hope that the results presented in this article stimulate further
research into the possibilities that field-induced adiabatic potentials
offer for the creation of new kinds of trapping potentials and
lower-dimensional geometries.  We have focussed in this paper on shell-like
and disc-shaped traps, but by working with atomic waveguides we also expect
that new tubular potentials could be formed.  These new kinds of potentials
can all be used to create new quantum-mechanical states of matter, and might
also be used in the study of weakly bound clusters or nano-particles.

This work was supported by the United Kingdom Engineering and Physical
Sciences Research Council. We would like to thank M.~Boshier, C.~Eberlein,
and E.~Hinds for discussions and comments.

\appendix*
\section{}

In this Appendix, we outline the derivation of Eq.\ (\ref{eqE}) for the
lifetime of the bubble states. The method is described in detail in
Ref.\ \cite{BanChi70} (see also \cite{Chi74}) so we can restrict
ourselves to indicating the main steps.

Consider the time-independent linear version of Eq.\ (\ref{GPE})
\begin{eqnarray}\label{GPE3}
 \varepsilon\phi_1&=&\left(-\frac 1 2 \frac{\partial^2}{\partial r^2}
 +\frac {r^2}2
 -\frac{\Delta}2\right)\phi_1+\Omega \phi_2, \nonumber \\
 \varepsilon\phi_2&=&\left(-\frac 1 2 \frac{\partial^2}{\partial r^2}
 -\frac {r^2} 2 +\frac{\Delta}2\right)\phi_2+\Omega \phi_1
\end{eqnarray}
with $\varepsilon$ the eigenenergy. We are interested in the structure
of the solutions at energies $\varepsilon > \Omega$ for which resonance
states are expected to appear. Far away from the potential crossing
at $r_c=\sqrt{\Delta}$ the eigenfunctions $\phi_{1,2}$ can be approximated 
as
(cf.\ Eq.\ (9) of \cite{BanChi70})
\begin{eqnarray}
\phi_1(r\ll \sqrt{\Delta})&\sim& k_1^{-1/2}\left[A_1(\infty)
\exp\left(i\int_{r_1}^r k_1 dr +i\pi/4\right)+\right. \nonumber \\
& & \left. A_1(-\infty)\exp\left(-i\int_{r_1}^r k_1 dr-i\pi/4\right) 
\right]
\nonumber \\
\phi_2(r\gg \sqrt{\Delta})&\sim& k_2^{-1/2}\left[A_2(\infty)\exp
\left(i\int_{r_2}^r k_2 dr-i\pi/4\right)+\right. \nonumber \\
& & \left. A_2(-\infty)\exp\left(-i\int_{r_1}^r k_1 dr-i\pi/4\right) 
\right]
\end{eqnarray}
with the classical momenta $k_1=(2\varepsilon -r^2+\Delta)^{1/2}$,
$k_2=(2\varepsilon+r^2-\Delta)^{1/2}$ and the turning points
$r_1=\sqrt{\Delta+2\varepsilon}$, $r_2=\sqrt{\Delta-2\varepsilon}$
(or $r_2=0$ if $\varepsilon>\Delta/2$). If we set, e.g., $A_2(-\infty)$
equal to unity then the other coefficients $A_i(\pm\infty)$ can be regarded
as scattering amplitudes for the interaction process inside the curve
crossing region. Resonance states are related to poles of the scattering
amplitudes in the complex energy plane. The determination of the
amplitudes proceeds in two steps. (i) By Taylor-expanding the harmonic
potentials to first order around the crossing at $r=r_c$, the interaction
between the states 1 and 2 is described as a linear
curve-crossing problem. Using the corresponding scattering matrix
the amplitudes $A_1(\infty)$ and $A_2(\infty)$ can be expressed as
linear functions of $A_1(-\infty)$ and $A_2(-\infty)$. The explicit
formulas (Eq.\ (19) of \cite{BanChi70}) involve the quantities
$\delta$ and $\Phi$ introduced in Sec.\ III.B. Note that the analytic
form of the scattering matrix is only known approximately (see, e.g.,
Sec.\ IV of \cite{ZhuNak93} for a comparison of different results), so that
quantitative calculations based on the analytic approach depend somewhat
on the expressions used. (ii) The wave function $\phi_1$ has to be 
subjected
to appropriate boundary conditions, i.e., $\phi_1(r=0)=0$. This immediately
yields the relation
\begin{equation}
A_1(\infty)/A_1(-\infty)=-\exp[2i\beta(\varepsilon)]
\end{equation}
where
$\beta(\varepsilon)=\int_0^{r_1}k_1(r)dr-\pi/4=\pi(2\varepsilon+\Delta-1)/4$.
Combining the results of (i) and (ii) and setting $A_2(-\infty)=1$ we
finally obtain (cf. Eq.\ (29) of \cite{BanChi70}, note the sign error)
\begin{eqnarray}
& & A_2(\infty)= \\ \nonumber
& & \frac{\cos\beta(E)+[e^{2\pi \delta(E)}-1]\cos\Phi(E)
e^{i[\beta(E)-\Phi(E)]}}{\cos\beta(E)+[e^{2\pi \delta(E)}-1]\cos\Phi(E)
e^{-i[\beta(E)-\Phi(E)]}}.
\end{eqnarray}
The poles of $A_2(\infty)$ are determined by condition (\ref{eqE}).

\end{document}